\documentclass[twocolumn,aps,prl,amsmath,amssymb,showpacs,superscriptaddress]{revtex4}

\usepackage{graphicx}
\usepackage{graphics}
\usepackage[latin1]{inputenc}
\usepackage{latexsym}
\usepackage{amsmath}
\usepackage{amssymb}

\newcommand{\be}{\begin{equation}}
\newcommand{\ee}{\end{equation}}
\newcommand{\eea}{\end{eqnarray}}
\newcommand{\bea}{\begin{eqnarray}}

\newcommand{\mean}[1]{\ensuremath{\langle{#1}\rangle}}

\newcommand{\BB}{\ensuremath{\mathcal{B}}}

\newcommand{\kommentar}[1]{}

\renewcommand{\vr}{\ensuremath{\varrho}}

\newcommand{\forget}[1]{}

\begin{document}

\title{Bell inequality tests of Four-Photon Six-Qubit Graph States}

%%%%%%%%%%%%%%%%%%%%%%%%%%%%%%%%%%%%%%%%%%%%%%%%%%%%%%%%%%%%%%%%%%%

\author{Wei-Bo Gao}
\affiliation{Hefei National Laboratory for Physical Sciences at
Microscale and Department of Modern Physics, University of Science
and Technology of China, Hefei, Anhui 230026, China}
\author{Xing-Can Yao}
\affiliation{Hefei National Laboratory for Physical Sciences at
Microscale and Department of Modern Physics, University of Science
and Technology of China, Hefei, Anhui 230026, China}
\author{Ping Xu}
\affiliation{Hefei National Laboratory for Physical Sciences at
Microscale and Department of Modern Physics, University of Science
and Technology of China, Hefei, Anhui 230026, China}
\author{He Lu}
\affiliation{Hefei National Laboratory for Physical Sciences at
Microscale and Department of Modern Physics, University of Science
and Technology of China, Hefei, Anhui 230026, China}
\author{Otfried G\"{u}hne}
\affiliation{Institut f\"{u}r Quantenoptik und Quanteninformation,
\"{O}sterreichische Akademie der Wissenschaften, Technikerstra{\ss}e
21A, A-6020 Innsbruck, Austria}
\affiliation{Institut f\"{u}r theoretische Physik,
Universit\"at Innsbruck, Technikerstra{\ss}e
25, A-6020 Innsbruck, Austria}
\author{Ad\'{a}n Cabello}
\affiliation{Departamento de F\'{\i}sica Aplicada II, Universidad de
Sevilla, E-41012 Sevilla, Spain}
\author{Chao-Yang Lu}
\affiliation{Hefei National Laboratory for Physical Sciences at
Microscale and Department of Modern Physics, University of Science
and Technology of China, Hefei, Anhui 230026, China}
\author{Tao Yang}
\affiliation{Hefei National Laboratory for Physical Sciences at
Microscale and Department of Modern Physics, University of Science
and Technology of China, Hefei, Anhui 230026, China}
\author{Zeng-Bing Chen}
\affiliation{Hefei National Laboratory for Physical Sciences at
Microscale and Department of Modern Physics, University of Science
and Technology of China, Hefei, Anhui 230026, China}
\author{Jian-Wei Pan}
\affiliation{Hefei National Laboratory for Physical Sciences at
Microscale and Department of Modern Physics, University of Science
and Technology of China, Hefei, Anhui 230026, China}
\affiliation{Physikalisches Institut, Ruprecht-Karls-Universit\"{a}t
Heidelberg, Philosophenweg 12, 69120 Heidelberg, Germany}

%%%%%%%%%%%%%%%%%%%%%%%%%%%%%%%%%%%%%%%%%%%%%%%%%%%%%%%%%%%%%%%%%%%

\date{\today}
%First version: July 2008
%This version: 11 March 2009

%%%%%%%%%%%%%%%%%%%%%%%%%%%%%%%%%%%%%%%%%%%%%%%%%%%%%%%%%%%%%%%%%%%

\begin{abstract}
We demonstrate experimentally a Y-shape graph state with photons'
polarization and spatial modes as qubits for the first time. Based
on the state and a linear-type graph state, we report on the
experimental realization of two different Bell inequality tests,
which represent higher violation than previous Bell tests.
\end{abstract}

%%%%%%%%%%%%%%%%%%%%%%%%%%%%%%%%%%%%%%%%%%%%%%%%%%%%%%%%%%%%%%%%%%%

%Entanglement measures, witnesses, and other characterizations
%42.50.Xa
%Optical tests of quantum theory

\maketitle

%%%%%%%%%%%%%%%%%%%%%%%%%%%%%%%%%%%%%%%%%%%%%%%%%%%%%%%%%%%%%%%%%%%

%%%%%%%%%%%%%%%%%%%%%%%%%%%%%%%%%%%%%%%%%%%%%%%%%%%%%5
\section{I. Introduction}%%%%%%%%%%%%%%%%%%%%%%%%%%%%%%%%%%%%%%%%%%%%%%%%%%%%%%%%

Graph states are basic resources for one-way quantum computation
\cite{RB01}, quantum error-correction \cite{SW02},
%quantum communication protocols \cite{Cleve},
and studying multiparticle entanglement \cite{HEB04}. Moreover, they
provide a test-bed to investigate quantum nonlocality, that is, the
inconsistency between local hidden variable (LHV) theories and
quantum mechanics \cite{GTHB05,otfried,avn,CGR08, mermin}.
Considerable efforts have been devoted to designing different Bell
inequalities for graph states with many particles. Here, the aim is
to find inequalities with a high quantum mechanical violation, as
this is related to the detection efficiency required to perform a
loophole-free testing of Bell inequality; moreover, the Bell
inequality with a higher violation is more robust against noise. In
these studies it has turned out that for many kinds of graph states,
the violation of local realism increases exponentially with the
number of particles \cite{mermin, otfried}. Experimental Bell tests
with four-qubit cluster or Greenberger-Horne-Zeilinger (GHZ) states,
which are examples of graph states, have been reported recently
\cite{walther,Kiesel,Matini1,Zhi}.

In this paper we report an experimental realization of a Y-shape
graph states Y$_6$, which are produced using the polarization and
the spatial modes of four photons. Such states are also called
hyper-entangled states and can be generated with good quality and a
high generation rate
\cite{Kwiat1,Kwiat2,Matini1,Matini2,Kai,bangzi,Gao}. Based on the
state and a linear-type one LC$_6$, we demonstrate two six-qubit
Bell tests, which remarkably represent higher violation than
previous experiments on Bell tests. In addition, we give a simple
theoretical proof that they give the same high violation of local
realism as the six-qubit GHZ state with the Mermin inequality
\cite{mermin}, but their violation can be more robust against
decoherence in principle.

%%%%%%%%%%%%%%%%%%%%%%%%%%%%%%%%%%%%%%%%%%%%%%%%%%%%%%%%%%%%%%%%%%%

%%%%%%%%%%%%%%%%%%%%%%%%%%%%%%%%%%%%%%%%%%%%%%%%%%%%%%%%%%%
\section{II. State preparation}

Let us first recall the notion of graph states. A graph state
$|G\rangle$ is specified by its stabilizer \cite{HEB04}, i.e., a
complete set of operators $g_i$ of which it is the unique joint
eigenstate, $g_i \, |G\rangle =|G\rangle$ for all $i$, where
\begin{equation}
\label{K} g_i = X_i \bigotimes_{j\in N(i)} Z_j.
\end{equation}
Here, $i$ is some vertex in a graph (see also Fig. 1a) and $N(i)$
denotes its neighborhood, that is, all vertices connected with
$i$. Furthermore, $X_i$ and $Z_j$ denote the usual Pauli operators
acting on qubits $i$ or $j$.

Below we demonstrate the creation of the desired state. The graph
corresponding to the Y-shape graph state is given in Fig.~1a (right)
and the experimental setup is shown in Fig.~1b. First, we use
spontaneous parametric down conversion \cite{K95, Z04} to create one
entangled photon pair $(\left| H \right\rangle _1 \left| H
\right\rangle _2 + \left| V \right\rangle _1 \left| V \right\rangle
_2 )/{{\sqrt 2 }}$ and two single photons $\left|  + \right\rangle =
(\left| H \right\rangle + \left| V \right\rangle )/{\sqrt 2 }$,
where $H$, $V$ denote horizontal and vertical polarization, and 1, 2
label the spatial modes of the photons. By using operations similar
to fusion-II gates between photons above \cite{BR05}, we generate a
state in
\begin{eqnarray}
\left| {LC_4 } \right\rangle = \frac{1}{2} \big[ \left| H
\right\rangle _1\left| H \right\rangle _3 (\left| H \right\rangle _2
\left| H \right\rangle _4 + \left| V \right\rangle _2 \left| V
\right\rangle _4 ) \notag
\\
+ \left| V \right\rangle _1 \left| V \right\rangle _3
(\left| H \right\rangle _2 \left| H \right\rangle _4
- \left| V \right\rangle
_2 \left| V \right\rangle _4 )\big],
\label{1}
\end{eqnarray}
which is equivalent to a  4-photon linear-type cluster state under
local unitary transformations \cite{yama}. Based on the state
$\left| {LC_4 } \right\rangle$, we apply two Hadamard (H) gates on
photons 2 and 4. Then, another two qubits in spatial modes are added
to construct the 6-qubit state. If a beam of photons enter a
polarizing beam splitter (PBS), the $H$-polarized one will follow
one path, while the $V$-polarized one will follow the other path.
Here we define the first path as the photon's $H'$ spatial mode, and
the latter one as its $V'$ spatial mode. After we place two PBSs in
the outputs of photons 1 and 4, the
 whole state will be converted to
 \begin{align}
\left| {Y_6 } \right\rangle & = \frac{1}{2} \big\{ \left| H
\right\rangle _1 \left| H \right\rangle _3 \left| H \right\rangle _2
\left| H \right\rangle _4 \left| H' \right\rangle _1 \left| H'
\right\rangle _4
 + \left| H \right\rangle _1 \left| H \right\rangle _3
\nonumber
\\
&  \left| V \right\rangle _2\left| V \right\rangle _4 \left| H'
\right\rangle _1 \left| V' \right\rangle _4
 + \left| V \right\rangle _1 \left| V \right\rangle _3 \left| H \right\rangle _2 \left| V \right\rangle _4 \left| V' \right\rangle _1 \left| V' \right\rangle _4
\notag \\
& + \left| V \right\rangle _1 \left| V \right\rangle _3 \left| V
\right\rangle _2 \left| H \right\rangle _4 \left| V' \right\rangle
_1 \left| H' \right\rangle _4 \big\}
\\
= \frac{1}{2}& \big\{ \left| 0 \right\rangle _1 \left| 0
\right\rangle _3 \left| 0 \right\rangle _2 \left| 0 \right\rangle _4
\left| 0 \right\rangle _5 \left| 0 \right\rangle _6
%\notag \\
+ \left| 0 \right\rangle _1 \left| 0 \right\rangle _3 \left| 1
\right\rangle _2 \left| 1 \right\rangle _4 \left| 0 \right\rangle _5
\left| 1 \right\rangle _6
\notag \\
+& \left| 1 \right\rangle _1 \left| 1 \right\rangle _3 \left| 0
\right\rangle _2 \left| 1 \right\rangle _4 \left| 1 \right\rangle _5
\left| 1 \right\rangle _6 + \left| 1 \right\rangle _1 \left| 1
\right\rangle _3 \left| 1 \right\rangle _2 \left| 0 \right\rangle _4
\left| 1 \right\rangle _5 \left| 0 \right\rangle _6 \! \big\}.
\nonumber
\end{align}
This is equivalent to a Y-shape 6-qubit graph state up to single
qubit unitary transformations.

In the above procedure, if we apply two H gates on photons 1 and 4
instead of 2 and 4, the state will be a linear-type graph state
\cite{gao22} (see Fig.~1a (right))
\begin{align}
\left| {LC_6 } \right\rangle &= \frac{1}{{\sqrt{8}}} \Big\{ \big[
\left| 0 \right\rangle _5 \left| 0 \right\rangle _1 + \left| 1
\right\rangle _5 \left| 1 \right\rangle _1 \big] \left| 0
\right\rangle _3\otimes \big[ \left| \tilde{0} \right\rangle _2
\left| 0 \right\rangle _4 \left| 0 \right\rangle _6 \nonumber\\
&+ \left|  \tilde{1}  \right\rangle _2 \left| 1 \right\rangle _4
\left| 1 \right\rangle _6 \big] + \big[ \left| 0 \right\rangle _5
\left| 0 \right\rangle _1 - \left| 1 \right\rangle _5 \left| 1
\right\rangle _1 \big] \left| 1 \right\rangle _3\notag \\
&\otimes \big[ \left|  \tilde{1}  \right\rangle _2 \left| 0
\right\rangle _4 \left| 0 \right\rangle _6  + \left|  \tilde{0}
\right\rangle _2 \left| 1 \right\rangle _4 \left| 1 \right\rangle _6
\big] \Big\}, \label{eq4}
\end{align}
where $|  \tilde{0}\rangle = (\left| 0 \right\rangle+\left|  1
\right\rangle)/{\sqrt{2}}$, and $| \tilde{1}\rangle = (\left| 0
\right\rangle - \left| 1 \right\rangle)/{\sqrt{2}}$. $\left| {LC_6 }
\right\rangle$ is equivalent to a 6-qubit linear-type graph state up
to single qubit unitary transformations.

%%%%%%%%%%%%%%%%%%%%%%%%%%%%%%%%%%%%%%%%%%%%%%%%%%%%%%%%%%%%%%%%%%%

\begin{figure}
\includegraphics[width=8.2cm]{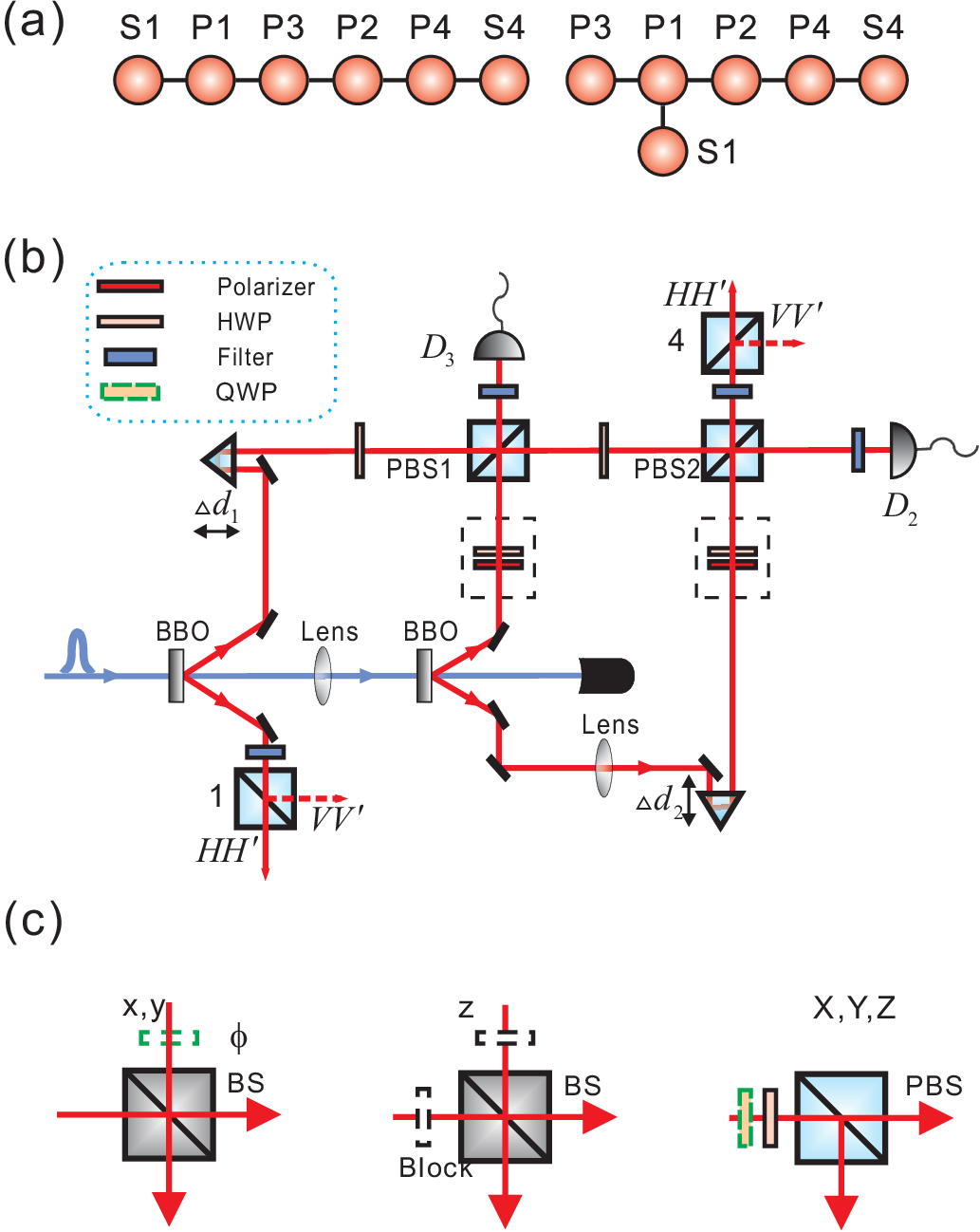}
\caption{(Color online) $\mathbf{a}$. The two graph states created.
$P_i$ represents (polarization) qubit $i$, and S1 and S4 represent
(spatial) qubits 5 and 6, respectively. $\mathbf{b}$. Scheme of the
experimental setup to generate the desired graph states. Femtosecond
laser pulses ($\approx$ 200 fs, 76 MHz, 788 nm) are converted to
ultraviolet and transmitted through two BBO crystals (2 mm), where
two photon pairs are generated. The observed two-fold coincident
count rate is about 2.6 $\times$ $10^{4}$/s. Two additional
polarizers are inserted into the arms of the second pair to prepare
two single photon states. $\mathbf{c}$. The measurement setups for
the desired observables. The first setup is for $x$ measurement of
spatial qubits when $\phi=0$, and for $y$ measurement of spatial
qubits when $\phi=90^\circ$. The second one is for $z$ measurement
of spatial qubits by using blocks in the two paths of the beam
splitter. The third one is for $X, Y, Z$ measurements of
polarization qubits by using half wave plates (HWPs), quarter wave
plates (QWPs), and PBSs.} \label{2}
\end{figure}

%%%%%%%%%%%%%%%%%%%%%%%%%%%%%%%%%%%%%%%%%%%%%%%%%%%%%%%%%%%%%%%%%%%

%%%%%%%%%%%%%%%%%%%%%%%%%%%%%%%%%%%%%%%%%%%%%%%%%%%%%%%%%%%%%%%%%%%

\begin{figure}
\includegraphics[width=7.8cm]{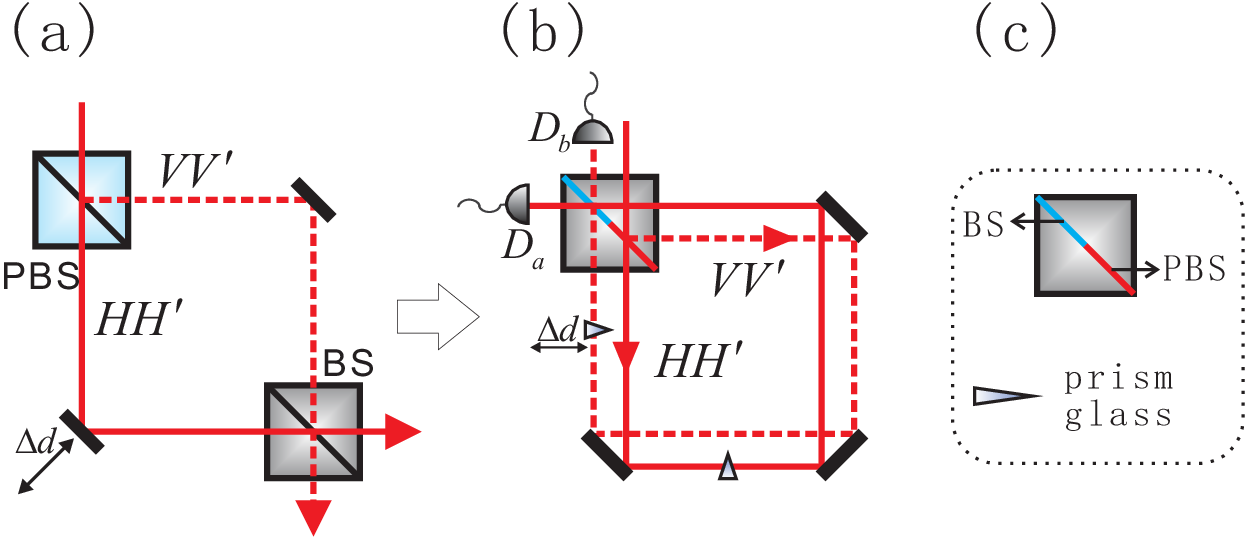}
\caption{(Color online) Apparatuses of constructing Sagnac-ring
interferometer in order to measure all the necessary observables of
spatial modes. $\mathbf{a}$. The original scheme of single photon
interferometer, which is easily affected by the environment and can
be stable for only several minutes. $\mathbf{b}$. Our single photon
interferometer in the Sagnac-ring model. Two special prism glasses
are inserted to change optical path delay in order to obtain the
desired phase. $\mathbf{c}$. A special crystal combining the
function of beam splitter and PBS.} \label{3}
\end{figure}

%%%%%%%%%%%%%%%%%%%%%%%%%%%%%%%%%%%%%%%%%%%%%%%%%%%%%%%%%%%%%%%%%%%

\section{III. results of the state fidelity}

In order to measure the states' fidelities and test the Bell
inequalities, we need to implement the desired local measurements.
The measurement setups are shown in Fig.~1c, which are similar to
Refs.~\cite{Kai, Matini1}. Here and in the following, $x$, $y$, $z$
refer to the Pauli matrices for the spatial modes, and $X$, $Y$, $Z$
refer to the Pauli matrices of the polarization modes. The
measurements of $x, y$ observables are implemented by overlapping
different modes of a photon on a beam splitter (BS), and the
measurement of $z$ observable is implemented by blocking one or the
other input path of the BS. The observables of polarization qubits
are measured by placing a combination of a quarter-wave plate, a
half-wave plate and a PBS in front of the single-photon detectors.
Although a photon's polarization and spatial information is read out
simultaneously, they are independent measurements and  have no
influence on each other.

The measurements of spatial modes require single photon
interferometers as shown in Fig.~2a. This interferometer is very
easily affected by its environment and can only be stable for a few
minutes. In our experiment, an ultra-stable Sagnac-ring technique
\cite{N07, A07} is applied to satisfy the required stability. First,
we design a crystal combining a PBS and a BS as shown in Fig.~2c.
Then, we construct the single photon interferometer in a Sagnac-ring
configuration (see Fig.~2b). The $H$-polarized component is
transmitted and propagated through the interferometer in the
counterclockwise direction, while the $V$-polarized component is
reflected and propagated through the interferometer in the clockwise
direction. Then, the interference happens when the two components
meet at the BS. Such interferometer can be stable for at least ten
hours \cite{Gao}.

%%%%%%%%%%%%%%%%%%%%%%%%%%%%%%%%%%%%%%%%%%%%%%%%%%%%%%%%%%%%%%

%%%%%%%%%%%%%%%%%%%%%%%%%%%%%%%%%%%%%%%%%%%%%%%%%%%%%%%%%
%In order to characterize the generated states
%further and to prove the genuine multipartite 6-qubit entanglement,
To estimate the fidelity of the prepared states, we consider an
observable $B$ with the property $ \left\langle \phi \right |B
\left| \phi \right\rangle \leq \left\langle {\phi } | {Y_6 }\rangle
\langle {Y_6 }|
 {\phi } \right\rangle  = F_{Y_6 }$
for any $|\phi\rangle.$ This means $\mean{B}_{\rm exp}$ is a lower
bound of the fidelity of the experimentally produced state
\cite{gtreview}. In the experiment, we have chosen the observable
$B$ in Ref. ~\cite{Y6remark} and find $\mean{B}_{\exp }  = 0.63 \pm
0.04$, clearly exceeding 1/2 and thus proving the genuine 6-qubit
entanglement of the state \cite{gtreview}. The fidelity of the
linear-type graph state is above $0.61 \pm 0.01$ \cite{gao22}, also
proving the genuine 6-qubit entanglement.

%%%%%%%%%%%%%%%%%%%%%%%%%%%%%%%%%%%%%%%%%%%%%%%%5

\section{IV. Results of Optimal Bell inequalities}

The optimal Bell inequality (i.e., the one having the highest
resistance to noise) involving only stabilizing observables for the
LC$_6$ state in the form of Eq.~(\ref{eq4}) is
\begin{equation}
\mean{\BB_{LC6}}
=
\mean{(\openone+g_5) g_1 (\openone+g_3)
(\openone+g_2) g_4 (\openone+g_6)}
\le 4,
\end{equation}
where $g_5=z_5 Z_1$, $g_1=x_5 X_1 Z_3$, $g_3=Z_1 X_3 Z_2$, $g_2=Z_3
X_2 Z_4$, $g_4=Z_2 X_4 x_6$ and $g_6=Z_4 z_6$ \cite{CGR08}. These
$g_i$ are stabilizing operators of the linear-type graph state, i.e.
the graph state is an eigenstate of all the $g_i$ with eigenvalue
$+1,$ as one can easily check. This writing of the Bell operator
is only a short-hand notation, and the required measurements for
the Bell test are the ones which arise {\it after} multiplying out
$\BB_{LC6}$ (see Table \ref{belllc6}).
As all the terms in the Bell operator are products of stabilizing
operators, the cluster state is an eigenstate of all these terms,
and the value for the ideal cluster state is the algebraic maximum
$\mean{\BB_{LC6}}=16.$

Similarly, the optimal stabilizer Bell inequality for
the Y$_6$ state is \cite{CGR08}
\begin{equation}
\mean{\BB_{Y6}}
=
\mean{
(\openone+g_3) g_1 (\openone+g_5) (\openone+g_2) g_4 (\openone+g_6)
}
\le 4,
\end{equation}
where now
$g_3 = Z_1 Z_3$,
$g_1 = X_1 X_3 X_2 x_5$,
$g_5 = Z_1 z_5$,
$g_2 = Z_1 Z_2 Z_4$,
$g_4 = Z_2 Z_4 z_6$,
and
$g_6 = Z_4 z_6.$
Again, the value for the pure Y$_6$ state is $\mean{\BB_{Y6}}=16.$

A remarkable feature of these Bell inequalities is that the LC$_6$
state and the Y$_6$ state violate local realism by a factor of four,
which is also the violation for the six-qubit GHZ state, if only
stabilizing elements are considered (the optimal Bell inequality is
then the Mermin inequality \cite{CGR08}). However, the LC$_6$ and
Y$_6$ state are more resistant to decoherence than the GHZ$_6$
\cite{DB04}. In fact, one can directly see that if decoherence acts
as a depolarizing channel on each qubit, the violation of the Mermin
inequality for the GHZ$_6$ state decreases faster than for the graph
states considered here. Namely, if noise like $\vr \mapsto p \vr +
(1-p) \openone/2$ is acting on each qubit separately, the value of
the Mermin inequality decreases with $p^6$, as the Mermin inequality
consists only of full correlation terms. In our Bell inequalities,
however, half of the terms contain the identity on one qubit (see
Tables I and II), which means that they decay only with $p^5,$ and
the total violation decreases like $(p^6+p^5)/2.$ This proves that
the non-locality vs.~decoherence ratio of GHZ states is not
universal: there are states with a similar violation which are more
robust against decoherence.

\begin{table}[t]
\begin{tabular}{ccccc}
\hline\hline Observable &Value &Observable
&Value\\
\hline
$xXZZXx$& $0.61\pm0.04$ & $-xXZZYy$ &  $0.60\pm0.04$&\\
$xXIYYx$& $0.63\pm0.04$ & $xXIYXy$ &  $0.62\pm0.04$&\\
$-yYZZXx$& $0.55\pm0.04$ & $yYZZYy$ &  $0.56\pm0.04$&\\
$-yYIYYx$& $0.65\pm0.03$ & $-yYIYXy $ &  $0.56\pm0.04$&\\
$xYYIXx$& $0.58\pm0.04$ & $-xYYIYy$ &  $0.63\pm0.04$&\\
$xYXXYx$& $0.58\pm0.04$ & $xYXXXy$ &  $0.60\pm0.04$&\\
$yXYIXx$& $0.55\pm0.04$ & $-yXYIYy$ &  $0.56\pm0.04$&\\
$yXXXYx$& $0.57\pm0.04$ & $yXXXXy$ &  $ 0.60\pm0.04$&\\
\hline\hline
\end{tabular}
\caption{ Experimental values of the observables on $\left| {LC_6 }
\right\rangle$ required in the test of the optimal Bell inequality.
Each experimental value is obtained by measuring in an average time
of 400 seconds and considers the Poissonian counting statistics of
the raw detection events for the experimental errors. The order of
the qubits is 5-1-3-2-4-6.} \label{belllc6}
\end{table}

\begin{table}[t]
\begin{tabular}{ccccc}
\hline\hline Observable &Value &Observable
&Value\\
\hline
$-XXIYxy$& $0.62\pm0.04$ & $XYIYyy$ &  $ 0.59\pm0.05$&\\
$YYIYxy$& $0.58\pm0.05$ & $YXIYyy$ &  $0.61\pm0.05$&\\
$XXIXxx$& $0.56\pm0.04$ & $-XYIXyx$&  $0.54\pm0.04$&\\
$-YYIXxx$& $0.61\pm0.04$ & $-YXIXyx$ &  $0.63\pm0.04$&\\
$-YXZXxy$& $0.57\pm0.05$ & $YYZXyy$ &  $0.62\pm0.04$&\\
$-XYZXxy$& $ 0.55\pm0.04$ & $-XXZXyy$ &  $0.58\pm0.04$&\\
$-YXZYxx$& $0.54\pm0.04$ & $YYZYyx$ &  $0.57\pm0.05$&\\
$-XYZYxx$& $0.59\pm0.04$ & $-XXZYyx$ &  $0.54\pm0.04$&\\
\hline\hline
\end{tabular}
\caption{Experimental values of all the observables on
$|Y_{6}\rangle$ for the optimal Bell inequality measurement. Each
experimental value is obtained by measuring in an average time of
400 seconds and propagated Poissonian statistics of the raw
detection events is also considered. The order of the qubits is
1-3-2-4-5-6.} \label{belly6}
\end{table}

The experimental results are given in Tables \ref{belllc6} and
\ref{belly6}. From these data we find
\begin{eqnarray}
\label{10} \mean{\BB_{LC6}}_{\rm exp}  &= &9.40 \pm 0.16,
\notag \\
\mean{\BB_{Y6}}_{\rm exp} &=& 9.30 \pm 0.17 ,
\end{eqnarray}
which violate the classical bound by 34 and 31 standard deviations.

Let us consider the ratio $\mathcal{D}$ between the quantum value of
the Bell operator and its bound in LHV theories.
%This is related to
%the detection efficiency required to perform a loophole-free testing
%of Bell inequality and if $\mathcal{D}$ increases, the required
%detection efficiency decreases \cite{CRV08}.
Experimentally, we have
\begin{eqnarray}
\label{11}
\mathcal{D}_{LC6}=\mean{\BB_{LC6}}_{\rm
exp}/\mean{\BB_{LC6}}_{\rm LHV}  &= &2.35 \pm 0.04,
\notag \\
\mathcal{D}_{Y6}=\mean{\BB_{Y6}}_{\rm exp}/\mean{\BB_{Y6}}_{\rm LHV}
&=& 2.33 \pm 0.04.
\end{eqnarray}
These are larger values compared to previous experiments with
similar Bell inequalities for four-qubit cluster states: there
values of $\mathcal{D}$ from 1.29 to 1.70 have been achieved
\cite{walther,Kiesel,Matini1}; using a Bell inequality with
non-stabilizer observables for the four-qubit GHZ state,
$\mathcal{D}=2.22$ has been reached \cite{Zhi}. To our knowledge,
these were the best values obtained so far. Therefore, despite
of having a lower fidelity than in the four-qubit experiments, we
find a higher violation of local realism, which demonstrates that
the amount of nonlocality can increase with the number of qubits.
This might help in designing loophole-free Bell inequality tests
\cite{CRV08}.

We would like to add that the generation of the  graph states
and the observation of the Bell inequality violations using
hyperentanglement implies that some of the qubits are carried
by the same photon, and therefore cannot be spatially seperated.
So our setup cannot be used  to close the locality loophole. However,
as the measurements on the polarization qubit and the spatial qubit
are independent, such experiments can be viewed as a test of the
Kochen-Specker theorem \cite{kirchmair, amselem} in order to refute
noncontextual hidden variable models.

%%%%%%%%%%%%%%%%%%%%%%%%%%%%%%%%%%%%%%%%%%%%%%%%%%%%%%%%%%%%%%%

%%%%%%%%%%%%%%%%%%%%%%%%%%%%%%%%%%%%%%%%%%%%%%%%%%%%%%%%%%%%%%%5
\section{V. Conclusion}

We have created a Y-shape four-photon six-qubit graph states
entangled in the photons' polarization and spatial modes and proved
its genuine six-qubit entanglement. Further, we have implemented two
multi-qubit Bell tests based on them, which show the highest
violation of Bell inequality so far. It is interesting to
investigate the relationship between decoherence and nonlocality
further. The aim is to characterize states, which show a high
violation of local realism, while being still robust against
decoherence.

%%%%%%%%%%%%%%%%%%%%%%%%%%%%%%%%%%%%%%%%%%%%%%%%%%%%%%%%%%%%%%%%%%%

\begin{acknowledgments}
This work is supported by the NNSF of China, the CAS, and the
National Fundamental Research Program (under Grant No.~2006CB921900).
OG acknowledges support from the FWF (START prize)
and the EU (OLAQUI, SCALA, and QICS). AC acknowledges support by the projects
No.~P06-FQM-02243 and No.~FIS2008-05596.
\end{acknowledgments}

%%%%%%%%%%%%%%%%%%%%%%%%%%% References %%%%%%%%%%%%%%%%%%%%%%%%%%%%


\begin{thebibliography}{99}


\bibitem{RB01}
R. Raussendorf and H.J. Briegel, Phys. Rev. Lett. {\bf 86}, 5188
(2001); M. van den Nest {\it et al.}, Phys. Rev. Lett. {\bf 97},
150504 (2006).

%%%%%%%%%%%%%%%%%%%%%%%%%%%%%%%%%%%%%%%%%%%%%%%%%%%%%%%%%%%%%%%%%%%%%

\bibitem{SW02}
D. Schlingemann and R.F. Werner, Phys. Rev.~A {\bf 65}, 012308
(2001); D. Schlingemann, Quantum Inf. Comput. {\bf 2}, 307 (2002).


%%%%%%%%%%%%%%%%%%%%%%%%%%%%%%%%%%%%%%%%%%%%%%%%%%%%%%%%%%%%%%%%%%%%%
% \bibitem{Cleve}
% R. Cleve, D. Gottesman, and H.-K. Lo,
% Phys. Rev. Lett. {\bf 83}, 648 (1999).


%%%%%%%%%%%%%%%%%%%%%%%%%%5
\bibitem{HEB04}
M. Hein, J. Eisert, and H.J. Briegel, Phys. Rev.~A {\bf 69}, 062311
(2004); M. Hein {\it et al.}, in {\em Quantum Computers, Algorithms
and Chaos}, edited by G. Casati, D.L. Shepelyansky, P. Zoller, and
G. Benenti (IOS Press, Amsterdam, 2006); quant-ph/0602096.


%%%%%%%%%%%%%%%%%%%%%%%%%%%%%%%%%%%%%%%%%%%%%%%%%%%%%%%%%%%%%%%%%%%%%

\bibitem{GTHB05}
V. Scarani {\it et al.}, Phys. Rev. A {\bf 71}, 042325 (2005); J.
Barrett {\it et al.}, Phys. Rev. A {\bf 75}, 012103 (2007); O.
G\"{u}hne and A. Cabello, Phys. Rev. A {\bf 77}, 032108 (2008).

\bibitem{avn}
A. Cabello and P. Moreno, Phys. Rev. Lett. {\bf 99}, 220402 (2007).

\bibitem{CGR08}
A. Cabello, O. G{\"u}hne, and D. Rodr\'{i}guez, Phys. Rev. A {\bf
77}, 062106 (2008).

\bibitem{mermin}
N. D. Mermin, Phys. Rev. Lett. {\bf 65}, 1838 (1990).

\bibitem{otfried}
O.~G\"uhne {\it et al.}, Phys. Rev. Lett. {\bf 95}, 120405 (2005);
G.~T\'oth, O. G\"uhne, and H.J. Briegel, Phys. Rev. A {\bf 73},
022303 (2006); L.-Y. Hsu, Phys. Rev. A {\bf 73}, 042308 (2006).




%%%%%%%%%%%%%%%%%%%%%%%%%%%%%%%%%%%%%%%%%%%%%%%%%%%%%%%%%%%%%%%%%%%%%

\bibitem{walther}

P. Walther, M. Aspelmeyer, K. J. Resch, and A. Zeilinger, Phys. Rev.
Lett. {\bf 95}, 020403 (2005).
\bibitem{Kiesel}
N. Kiesel {\it et al.}, Phys. Rev. Lett. {\bf 95}, 210502 (2005).


\bibitem{Matini1} G. Vallone {\it et al.}, Phys. Rev. Lett. {\bf 98}, 180502 (2007).

\bibitem{Zhi} Z. Zhao {\it et al.}, Phys. Rev. Lett. {\bf 91}, 180401 (2003).
%%%%%%%%%%%%%%%%%%%%%%%%%%%%%%%%%%%%%%%%%%%%%%%%%%%%%%%%%%%%%%%%%

\bibitem{Kwiat1}
P. G. Kwiat, J. Mod. Opt. {\bf 44}, 2173 (1997).

\bibitem{Kwiat2}
J. T. Barreiro {\it et al.}, Phys. Rev. Lett. {\bf 95}, 260501
(2005).

\bibitem{Matini2}
G. Vallone {\it et al.}, Phys. Rev. Lett. {\bf 100}, 160502 (2008).


\bibitem{Kai}
K. Chen {\it et al.}, Phys. Rev. Lett. {\bf 99}, 120503 (2007).

\bibitem{bangzi}
H. S. Park {\it et al.}, Optics Express {\bf 15}, 17960 (2007).

\bibitem{Gao}
W.-B. Gao {\it et al.}, Nature Physics, {\bf 6}, 331 (2010).


%%%%%%%%%%%%%%%%%%%%%%%%%%%%%%%%%%%%%%%%
%%%%%%%%%%%%%%%%%%%%%%%%%%%%%%%%%%%%

\bibitem{K95}
P. G. Kwiat {\em et al.}, Phys. Rev. Lett. {\bf 75}, 4337 (1995).

\bibitem{Z04}
Z. Zhao {\em et al.}, Nature (London) {\bf 430}, 54 (2004).

\bibitem{BR05}
D. E. Browne and T. Rudolph, Phys. Rev. Lett. {\bf 95}, 010501
(2005).

\bibitem{yama}
Y. Tokunaga {\it et al.}, Phys. Rev. Lett. {\bf 100}, 210501 (2008).

\bibitem{gao22} W.-B. Gao {\it et al.}, Phys. Rev. Lett. {\bf 104}, 020501 (2010).

\bibitem{N07}
T. Nagata {\em et al.},
%
Science {\bf 316}, 726 (2007).

\bibitem{A07}
M. P. Almeida {\em et al.},
%
Science {\bf 319}, 579 (2007).

\bibitem{gtreview}
O. G\"uhne and  G. T\'oth, Phys. Rep. {\bf 474}, 1 (2009).

\bibitem{Y6remark}
See the online material of C.-Y. Lu {\em et al.}, Phys. Rev. Lett.
{\bf 102}, 030502 (2009).

\bibitem{DB04}
W. D\"ur and H.-J. Briegel, Phys. Rev. Lett. {\bf 92}, 180403
(2004).

\bibitem{CRV08}
A. Cabello, D. Rodr\'{\i}guez, and I. Villanueva,
%``Necessary and sufficient detection efficiency for the Mermin
%inequalities'',
Phys. Rev. Lett. {\bf 101}, 120402 (2008).

\bibitem{kirchmair}
G. Kirchmair {\em et al.} Nature {\bf 460}, 494 (2009).

\bibitem{amselem}
E. Amselem {\em et al.} Phys. Rev. Lett. {\bf 103}, 160405  (2009).

\end{thebibliography}
\end{document}